# Lorenz vs Lorentz

The controversy examined from the original texts

María Florentina Mendoza Urán
ORCID 0009-0007-7882-6418

---

**Abstract**

The attribution of the condition $\nabla \cdot A + (1/c^2)\partial\Phi/\partial t = 0$ has been under discussion since 1938 as a question of priority between Ludvig Lorenz and Hendrik Lorentz. This work argues that the problem is of a different nature, and that the discussion does not close because it has never been specified what is being attributed. Drawing on the three versions of the 1867 paper and on six further primary sources, it is documented that three separable acts have been designated under a single name: the obtaining of the relation between potentials, its imposition as a restriction on the admissible potentials, and its denomination as a gauge. It is further documented that in none of the three versions of the 1867 paper does the word potential appear, that the magnetic field does not figure in it as a quantity of the theory, and that the vocabulary in which that paper is described today enters the literature after it was written. It is concluded that the disjunction between the two denominations, posed without specifying which of the three acts is being attributed, is ill formed.

## 1. Introduction

In the paper to which a condition on the electromagnetic potentials is attributed, the word potential does not appear.

It does not appear in the Danish original, nor in the German translation its author sent to Poggendorff, nor in the English one made from the German.[1] Neither does the magnetic field figure as a quantity of the theory: the root «magnet» occurs three times in the twenty pages of the original, two of them in the opening paragraph, within a general enumeration of the forces which the science of the century had been connecting, and the third as a unit of measurement.[2]

What there is are four letters, $\bar{\Omega}$, α, β and γ. The first its author calls a new function; the other three, designations, and he introduces them saying that it is for brevity.[3]

The equation those letters compose is the one that appears today in the textbooks under two names. Some call it the Lorenz condition or gauge, after the Danish physicist Ludvig Valentin Lorenz (1829-1891). Others call it the Lorentz condition or gauge, after the Dutch physicist Hendrik Antoon Lorentz (1853-1928). The coexistence of the two spellings is not a typographical oversight: it answers to a discussion that has been open for more than eighty years.

The error of attribution was pointed out by Alfred O'Rahilly in 1938, by Edmund Taylor Whittaker in 1951 and by Jean Van Bladel in 1991.[4] The correction consolidated from 1998 onwards, when John David Jackson adopted the spelling Lorenz in the third edition of Classical Electrodynamics, and it was documented in detail in the 2001 paper by Jackson and Lev Borisovich Okun.[5] The dominant editorial practice has been shifting towards writing Lorenz.

And yet the discussion has not stopped. In 2008, Bozhidar Iliev counts the use of both spellings and concludes that the incorrect one still predominated.[6] Between 2008 and 2010, H. C. Potter and C. W. Wong return to the 1867 paper with incompatible readings of the same step.[7] In 2023, Jed Z. Buchwald argues that what Lorenz did is not a gauge transformation, and proposes that the name should revert to the Dutchman.[8] A settled question of priority does not generate that kind of literature.

This work argues that the discussion persists because it has not been specified what is being attributed. Under the single name of gauge condition, three distinct, separable and datable acts have been designated: the obtaining of the relation between potentials, its imposition as a restriction on the admissible potentials, and its denomination as a gauge. The first corresponds to Lorenz in 1867; the second, to Emil Wiechert in 1900 and to Hendrik Lorentz in 1904; the third, to Walter Heitler, between 1936 and 1954.

It further argues that this indeterminacy has rested in part on an editorial fact. The 1867 paper was written and first published in Danish, and the subsequent discussion has been conducted on its translations and, above all, on the descriptions the secondary literature gives of it. When one goes to the originals, elements appear that those descriptions do not record.

Section 2 documents which is the original and which the translations, and which version each author cites. Section 3 documents the relation as it stands in print. Section 4 examines in what terms Lorenz declares his own steps. Section 5 documents what his paper does not contain. Sections 6 and 7 document the other two acts. Section 8 examines why Lorenz's name disappeared from the literature between 1867 and 1900, and when the vocabulary of potentials applied to his paper enters it. Section 9 gathers the conclusions.

## 2. The original and the translations

The paper had three versions in a single year and a fourth thirty-one years later.

The first was Danish: «Om Identiteten af Lyssvingninger og elektriske Strømme», presented to the Royal Danish Academy of Sciences and Letters and published in 1867 in the Oversigt over det Kongelige Danske Videnskabernes Selskabs Forhandlinger.[9] The second, German, in the Annalen der Physik und Chemie.[10] The third, English, in the Philosophical Magazine.[11] The fourth, French, did not appear until 1898, in the first volume of the Oeuvres Scientifiques edited by Herman Valentiner.[12]

The order and the dependence between them do not have to be inferred: they are declared. The heading of the German version reads «Aus dem Oversigt over det K. Danske Vidensk. Selsk. Forhandl. 1867, No. 1, vom Hrn. Verfasser übermittelt», that is, from the Danish Oversigt, transmitted by the author himself.[13] And the heading of the English one reads «Translated from Poggendorff's Annalen, June 1867».[14] The English version is, therefore, a translation of a translation.

Helge Kragh, author of the first scientific biography of Lorenz, confirms this and adds the detail: in order to make his work accessible to an international audience, Lorenz sent a copy to Johann Christian Poggendorff, editor of the Annalen, where it appeared in German translation that same year; and the German version served as the basis for the English translation.[15]

The literature of the controversy cites, in general, the translated versions. Jackson reproduces the equation and locates it by the Philosophical Magazine pagination.[16] Potter translates into modern vector notation starting from the same version and keeping its equation numbering.[17] Wong likewise cites the English version.[18] Kragh constitutes the declared exception, and he himself notes that Danish material by Lorenz from that same year has been overlooked.[19]

The collation of the three versions, carried out for this work, yields no discrepancies in the passages that matter here. The Danish «antage» is «anzunehmen» and «taking»; «transformeres» is «verwandelt» and «transformed»; the «ubestemte Konstant» is «unbestimmte Größe» and «indeterminate magnitude»; the «Betegnelse» are «Bezeichnungen» and «designations». Where the Danish says derive, the other two say derive; where it says transform, they say transform. The translations are faithful.

The problem, therefore, does not lie in the chain of translation. It lies in the distance between what the three versions say and what the literature says they say.

## 3. The relation, as it stands in print

The symbols are defined by Lorenz himself. On page 290 of the English version he introduces a new function, $\bar{\Omega}$, which is the scalar potential of Kirchhoff's theory with $e'(t - r/a)$ written in place of $e'$. On page 291 he introduces α, β and γ as abbreviations for three integrals carrying inside the current density evaluated at the instant $t - r/a$. The formula of introduction is, in this second case, that it is for brevity.[20]

On page 294, after writing that there is some reason for taking $a = c/\sqrt{2}$ and that on substituting that value the equations assume a very simple form and lead to exactly the same differential equations as those he had formerly deduced for the vibrations of light, with the addition of only a single member, Lorenz writes: «by partial integration and introduction of the designations α, β, γ we obtain», and then

$$d\bar{\Omega}/dt = -2\ (d\alpha/dx + d\beta/dy + d\gamma/dz)$$

The expression appears unnumbered, without denomination, and within a paragraph of running text, as an intermediate step in a derivation. The following line continues: «Moreover from (5)», with the wave equation for α and the indication that the same holds for β and γ; and then, substituting those values in equations (A) after differentiating them with respect to t, and with $c = a\sqrt{2}$, he obtains the three numbered equations (8).[21]

The factors 2 and 4 that appear in the paper have their origin, according to Jackson, in a definition of electric current in terms of moving charges that was later abandoned.[22] Potter notes that Lorenz sets the speed of light in vacuum equal to $\sqrt{2}/2$ times the value measured by Weber for the speed of magnetic induction.[23]

The relation appears in the paper in two directions. In the first part, Lorenz obtains it: he starts from his retarded integrals and arrives at it after an integration by parts. In the final part he travels the inverse road, from the differential equations of the fields towards the integrals, and writes «If, in accordance with the earlier notation, we put» followed by the same relation.[24]

No author consulted denies that the relation is in the paper.

## 4. In what terms Lorenz declares what he does

The secondary literature employs different verbs when describing the 1867 paper, and the choice of verb decides whether the wave equations are in it a result or a starting point. It is therefore worth seeing in what terms its author describes them.

Lorenz declares his assumptions in writing. On page 29 of the Danish original he writes that Kirchhoff's equations were obtained in a purely empirical manner, that for that reason they are not necessarily the exact expression of the true law, and that it will always be permissible to add members to them or to give them another form so long as what has been established by experiment is not affected; and he announces that they will begin by considering the two members on the right-hand side of equations (1) as the first members of a series.[25] On page 32 he calls the velocity of propagation «denne ubestemte Konstant», this indeterminate constant.[26] On page 33 he expressly renounces the search for a physical hypothesis, because he has found that it can be done in several ways and for that reason the method loses all its value.[27]

On page 34 he writes «Vi have altsaa Grund til at antage $a = c/\sqrt{2}$», we have therefore reason to suppose $a = c/\sqrt{2}$.[28] The verb is «antage», to suppose or to assume, not «udlede», to derive. The value comes from Weber's determination divided by the square root of two, and Lorenz justifies it by its agreement with the measurements of the velocity of light.

On the passage from the integral form to the differential one he employs, on page 33, the verb «transformeres», are transformed.[29] The English version renders it «are transformed». He describes, then, his own operation as a change of form.

The paper contains four explicit references to earlier works of his own. On the theorem that permits the passage from the retarded integrals to the differential equations, he writes that the proof is to be found in his paper in volume 58 of Crelle's Journal, that is, in a work of 1861 on elasticity and not on electromagnetism.[30] On page 36 he writes that his differential equations for the components of the electric current now agree entirely with those he had formerly found for the components of light, referring to volume 121 of the Annalen, save for the last member, into which the electrical conductivity enters.[31] On

page 39 he writes that the boundary conditions are just the same as those he had already found for the components of light, referring to volume 118, and adds that for that calculation they need include no other assumptions than precisely those which the theory of light itself gives them.[32] And the wave solution he introduces with the formula «Disse Ligninger tilfredsstilles til Exempel ved», these equations are satisfied, for instance, by, after which he writes a damped plane wave and checks that it fits.[33]

On page 42, closing the inverse road, he writes «ere vi altsaa komne tilbage til den første Ligning (A) og de to andre udledes heraf ved Analogi»: we have, then, returned to the first equation (A), and the other two are derived from it by analogy. And he concludes that this result affords a new proof of the identity of the vibrations of light and the electric currents, since it is now seen that not only can the laws of light be derived from those of the electric currents, but that the converse road may also be travelled, adding precisely the same boundary conditions that the theory of light requires.[34]

The language of the literature agrees, in general, with that of the author. Jackson writes that Lorenz begins with the quasi-static potentials, with the vector potential in the Kirchhoff-Weber form, and proceeds toward the differential equations for the fields; and that the imposition of the relation leads to the simple wave equations whose solutions, as he had proved in 1861, are the standard retarded potentials.[35] Wong opens his abstract with Lorenz postulating that both potentials are retarded.[36] Potter notes that Lorenz considers the integral definitions to be Green's function solutions of his wave equations: he considers them already such.[37]

The heart of the matter is that the retarded potential is the Green's function solution of the wave equation with sources. They are not two distinct propositions from which one follows from the other: they are the same proposition written in integral and in differential form. The passage from one to the other is correct calculation, but not ampliative. And the connection between the two forms is not established in 1867: it comes from the 1861 work on elasticity, to which Lorenz refers.

## 5. What the paper does not contain

The word potential does not appear in any of the three versions. Nor does «gauge», nor «Eichung», nor «Green», nor «Maxwell», nor «dielectric», nor «displacement current».

And the magnetic field does not figure as a quantity of the theory. In the twenty pages of the Danish original the root «magnet» appears three times. Two are on page 26, in the opening paragraph, and belong to the general exposition of the problem: the enumeration of the connections found between the forces — between electricity and magnetism, heat, light, molecular and chemical forces— and the mention that Ampère theoretically explained the kinship between electricity and magnetism. The third is on page 37 and is «efter magnetisk Maal», according to magnetic measure, referring to the units in which Weber gave the conductivity of copper. Throughout the setting out, the development and the conclusion, the term does not appear again.[38]

This is not an isolated finding. Potter puts it thus: in his 1867 paper Lorenz never mentions the magnetic field and, therefore, never develops electromagnetic field equations.[39] Wong repeats it.[40] And Kragh adds that the permittivity of the vacuum had no place in Lorenz's theory, that light in it is conduction current with no analogue of Maxwell's displacement current, and that in a certain sense the theory was not electromagnetic, since it does not operate with the magnetic field.[41]

None of this contradicts Jackson's statement that Lorenz proceeds to derive the Ampère-Maxwell equation relating the curl of the magnetic field to the sum of the displacement current and the conduction current density.[42] Equations (8) have, in their structure, that form. Jackson describes the form of the equations; Kragh and Potter, the vocabulary and the scope of the theory. Both descriptions rest on the same text.

The consequence for attribution is the following. The relation now called a gauge condition appears in a paper whose author does not operate with the magnetic field, does not have the permittivity of the vacuum, does not employ the word potential and does not present what he is doing as a field theory. To attribute a gauge to him —that is, a condition on the potentials of a field theory, chosen among alternatives

— is to attribute to him an object his framework does not contain. What it does contain, in print and verifiable, is the relation.

## 6. The imposition: Wiechert, 1900; Lorentz, 1904

In 1888, George Francis FitzGerald arrives at the same relation by a different route and for an explicit physical motive. He was attempting to introduce a finite speed of propagation into his mechanical wheel-and-band model of the ether, and the instantaneous character of Maxwell's scalar potential was not compatible with it. On realising that this instantaneity was a consequence of $\nabla\cdot A = 0$, he proposed the relation and obtained the wave equations for both potentials.[43]

Thirty-three years after Lorenz, the first text in which the relation appears as a disposition over a quantity previously declared indeterminate is that of Emil Johann Wiechert, published in the volume commemorating the twenty-fifth anniversary of Lorentz's doctorate.[44] Wiechert starts from optics in the free ether, introduces two reciprocal transverse vector fields with vanishing divergences, expresses the magnetic field in terms of the vector potential Γ, and writes:

> *«Damit wird Γ noch nicht bestimmt; vor allem kommt in Betracht, dass der Werth von ∂Γx/∂x + ∂Γy/∂y + ∂Γz/∂z willkürlich bleibt; eine passende Verfügung behalten wir uns vor.»*

That is: with this Γ is not yet determined; above all it must be considered that the value of its divergence remains arbitrary; we reserve to ourselves a suitable disposition. And on the following page:[45]

> *«Über die Unbestimmtheit in Γ verfügend setzen wir nun: ∂Φ/∂t + V(∂Γx/∂x + ∂Γy/∂y + ∂Γz/∂z) = 0.»*

Disposing of the indeterminacy in Γ, we now set. The difference from the 1867 text lies not in the resulting equation, but in the status given to it. Lorenz obtains it from his integrals; Wiechert first declares that there is an indeterminate quantity and then disposes of it.

On that same page Wiechert adds an observation worth recording: that for Maxwell, Γ was not a mere mathematical auxiliary quantity but a function of the state with a meaning of its own, and that for that reason he had to leave the value of its divergence, though —and here he quotes in English within the German— «not related to any physical phenomenon», indeterminate.[46] Wiechert thus attributes to Maxwell the awareness of the indeterminacy.

Jackson further records that, on adding sources and establishing the retarded solutions, Wiechert cites other authors but not Lorenz.[47]

Thirty-seven years after Lorenz and four after Wiechert, Hendrik Antoon Lorentz writes the relation on page 157 of his article in the Encyklopädie der mathematischen Wissenschaften, in the form $\mathrm{div}\, a = -(1/c)\dot{\phi}$, and calls it a Relation.[48] His gesture does not coincide with Wiechert's. In that passage he points out that, although in every electromagnetic phenomenon the fields are determinate functions of the coordinates and of time, the potentials remain partly indeterminate; he says that he removes that inconvenience by subjecting them to the additional condition; he writes the general transformation relating any admissible pair to the first by means of a scalar function; and he states that this function can be determined so that the condition is satisfied.[49]

Wiechert declares an indeterminacy and disposes of it. Lorentz sets out the whole set of the alternatives and chooses among them.

In the footnotes of that page Lorentz cites Liénard and Wiechert.[50] And thirteen pages earlier, in placing his theory, he writes that it carries forward, in a new form only possible after Maxwell's work, an idea already pursued by Riemann, C. Neumann and Betti, whom he cites with their references.[51] Lorenz does not appear.

Kirk T. McDonald has pointed out the consequence: the general gauge transformation was mentioned by Lorentz in 1904 on that page, together with a statement of his preference, without attribution, so that this gauge came to be known to many as the Lorentz gauge.[52]

## 7. The name: Heitler

Neither Lorenz nor Lorentz named the relation.

The word itself is later than all three. Hermann Weyl used Eichinvarianz in 1919 with the sense of scale invariance, within an attempt to unify gravitation and electromagnetism; the term «gauge» in English, with the modern sense, appears in 1929.[53] The phase symmetry that gives the concept its present content was discovered by Vladimir Fock in 1926.[54]

According to Jackson and Okun, Walter Heitler introduced «Lorentz relation» in the first edition of The Quantum Theory of Radiation, 1936, and «Lorentz gauge» in the third, 1954.[55] The second edition, of 1944, allows an intermediate state to be dated. The running head of its page 3 is «POTENTIALS. GAUGE INVARIANCE», and there, after equation (10), Heitler writes:

> *«(10) represents a relation between the potentials and is called the Lorentz relation.»*

In italics in the original. On the same page and on the next he names it twice more as «the Lorentz condition (10)». And on that same page 3 he writes, in quotation marks, that the invariance under that transformation is called «gauge invariance».[56]

That is: by 1944 there are already in the same book, and within a few lines of one another, the Lorentz relation, the Lorentz condition and gauge invariance; but not the Lorentz gauge. The two terms are joined later.

The relation is written, then, in 1867; it is imposed in 1900 and 1904; and it receives a name between 1936 and 1954. That name refers to the language and the conceptual framework of quantum field theory, not to the electrodynamics in which the relation was born.

## 8. How a name is lost and how a vocabulary enters

Between 1867 and 1900 Lorenz's name comes away from the relation. The documented explanations are several and do not exclude one another.

The first is Maxwell's criticism. In a paper of 1868, describing Lorenz's, he writes that it shows that, on Weber's theory, periodic electric disturbances would be propagated with a velocity equal to that of light; and that the propagation of attraction through space forms part of this hypothesis also, though the medium is not explicitly recognised. And then:[57]

> *«From the assumptions of both these papers we may draw the conclusions, first, that action and reaction are not always equal and opposite, and second, that apparatus may be constructed to generate any amount of work from its resources.»*

Maxwell demonstrates this immediately afterwards with two oppositely charged bodies joined by a rigid rod.

It is worth noticing the description. Maxwell attributes to Lorenz a medium his text does not recognise, and places his paper on Weber's theory. He does not attribute potentials to him.

Jackson and Okun call the objection devastating and point to it as an important contributing factor in the neglect.[58] Kirk T. McDonald holds that it was not valid.[59] And Kragh documents that the first critical reference to Lorenz's theory came not from Radicke's review in Bonn, but from Maxwell in Cambridge.[60]

The second explanation points to Lorenz himself. Kragh argues that his phenomenological attitude and his indifference towards Maxwellian theory were the principal reasons why his mature works exerted little influence; and he documents, from unpublished notes, that Lorenz had studied Maxwell's theory and was aware of his critical remarks, and that he never referred to it.[61]

The third is of a material order. Jackson writes that Lorenz lost out to the homophonous Dutchman, and to Wiechert, in good part because he was a Dane who published an appreciable part of his work only in Danish; that he died in 1891, just when Lorentz was most productive; and that by around 1900 his name had virtually disappeared from the literature.[62] Nevels and Shin place in 1892, one year after that death, the moment when Lorentz's papers on the retarded potential associate his name with the gauge.[63]

It remains to establish when the vocabulary of potentials applied to the 1867 paper enters. It is not introduced by Maxwell, who in 1868 describes it on Weber's theory. It appears in 1884, in Heinrich Hertz:[64]

> *«Riemann in 1858 and Lorenz in 1867, with a view to associating optical and electrical phenomena with one another, postulated the same or quite similar laws for the propagation of the potentials.»*

What happens between 1867 and 1884 is not that the equation changes: what changes is what its symbols mean. In Lorenz, α, β and γ are abbreviations for three integrals, introduced for brevity and without a name; and $\bar{\Omega}$ is a new function. They are not quantities of the theory, but short ways of writing what he already had. In Hertz, those same integrals are the vector potential: a quantity with a standing of its own, the one he has been handling since Neumann and since Helmholtz. The formula is the same; the status of what it names is not.

The context of that sentence matters. Hertz writes it immediately after showing that the vector-potentials are propagated with finite velocity and according to the same laws as the vibrations of light. His starting point, like Helmholtz's, is Neumann's vector potential: in a footnote of the same paper he defines the usual electromagnetics as that which regards the forces deduced from Neumann's laws of the potential as exactly applicable.[65] And the recognition has a formal basis: the integral Lorenz calls α, β, γ has the same form as Neumann's vector potential, with the time retarded inside. What the formula does not say is where it comes from. Neumann's is defined in order to calculate forces; Lorenz's is what remains after an integration by parts, and its author calls it an abbreviation. Hertz reads the 1867 paper from within his own framework, and recognises in it his own. From that moment on, Lorenz's paper is described in a vocabulary that is not his.

In 1889, in his Heidelberg lecture, Hertz names them again, but in another manner: he writes that Maxwell's road had suggested to Riemann and Lorenz speculations of a similar nature, although not so fruitful in results.[66] There he does not employ the word potential in reference to them.

## 9. Conclusions

The three acts that have been designated by a single name are separable and dated.

The obtaining of the relation between potentials corresponds to Lorenz, in 1867. It appears unnumbered and without denomination, as an intermediate step in a derivation, and in the same paper it is travelled in both directions. No author consulted denies it.

The imposition corresponds to Wiechert in 1900 and to Lorentz in 1904, and in each it takes a different form. Wiechert declares that the divergence of the vector potential remains arbitrary and reserves the disposal of it to himself. Lorentz sets out the general transformation, acknowledges that other admissible pairs exist, and chooses. Neither of the two operations appears in the 1867 paper.

The denomination corresponds to Heitler, between 1936 and 1954. Neither Lorenz nor Lorentz named the relation, and the word «gauge» in its modern sense dates from 1929.

From this it follows that the disjunction «Lorenz or Lorentz», posed without specifying which of the three acts is being attributed, is ill formed. If what is attributed is the relation, it corresponds to Lorenz, with the antecedents of Kirchhoff, Riemann and Helmholtz, and with FitzGerald arriving at it independently in 1888. If what is attributed is the act of imposing it, it corresponds to Wiechert and to Lorentz, in that order. And if what is attributed is the name, it corresponds to Heitler.

A conclusion about the object itself also follows. The 1867 paper does not operate with the magnetic field, does not have the permittivity of the vacuum, does not employ the word potential and does not present itself as a field theory. The relation it contains is therefore prior to the framework in which the word

«gauge» has a sense. To attribute a gauge to that paper is not merely a question of a name: it is to apply to it a category its author did not handle.

And there follows, lastly, an observation of method. This discussion has been conducted for eighty years on translations and on descriptions of a text written in Danish. The terms in which Lorenz describes his own operations —to suppose, to transform, to return, to agree— are in the original, and they are what allow one to establish what he did and what he did not. The attribution of an idea depends on the verb with which what its author did is described; and that verb, when one works on descriptions made by others, ceases to be his own.

## Notes

**1.** See notes 9, 10 and 11 for the three references. The count has been carried out on the complete text of the Danish and the English versions, and on the transcription of the German one.

**2.** Lorenz 1867 (Danish version), pp. 26 and 37. Count over the complete text, pp. 26-45.

**3.** Lorenz 1867 (English version), p. 290: «Let a new function $\bar{\Omega}$ be defined»; and p. 291: «where, for brevity's sake, we put».

**4.** Alfred O'Rahilly, Electromagnetics (Longman, Green & Co. and Cork University Press, 1938); reprinted with corrections, Electromagnetic Theory (Dover, New York, 1965). E. T. Whittaker, A History of the Theories of Aether and Electricity, vol. I, revised ed. (Nelson, London, 1951), p. 268. Jean Van Bladel, «Lorenz or Lorentz?», IEEE Antennas and Propagation Magazine 33, no. 2, 69 (1991).

**5.** J. D. Jackson, Classical Electrodynamics, 3rd ed. (Wiley, New York, 1998), sec. 6.3. J. D. Jackson and L. B. Okun, «Historical roots of gauge invariance», Reviews of Modern Physics 73, 663-680 (2001).

**6.** B. Z. Iliev, «The Lorenz gauge is named in honour of Ludwig Valentin Lorenz!», arXiv:0803.0047 (2008).

**7.** H. C. Potter, arXiv:0810.1172 (2008), arXiv:0811.2123 (2008) and arXiv:0903.4083 (2009); C. W. Wong, arXiv:1012.4128 (2010).

**8.** Jed Z. Buchwald, «Gauging Potentials: Maxwell, Lorenz, Lorentz and Others on Linking the Electric Scalar and Vector Potentials», in The Richness of the History of Mathematics, eds. Karine Chemla et al., Archimedes 66 (Springer, 2023), pp. 341-364.

**9.** L. V. Lorenz, «Om Identiteten af Lyssvingninger og elektriske Strømme», Oversigt over det Kongelige Danske Videnskabernes Selskabs Forhandlinger, 1867, no. 1, 26-45. Printed in 1868.

**10.** L. V. Lorenz, «Ueber die Identität der Schwingungen des Lichts mit den elektrischen Strömen», Annalen der Physik und Chemie 131, 243-263 (1867). DOI 10.1002/andp.18672070606.

**11.** L. V. Lorenz, «XXXVIII. On the Identity of the Vibrations of Light with Electrical Currents», Philosophical Magazine, series 4, 34, no. 230, 287-301 (1867). DOI 10.1080/14786446708639882.

**12.** Herman Valentiner (ed.), Oeuvres scientifiques de L. Lorenz, revues et annotées, 2 vols. (Copenhagen, 1898-1904), published at the expense of the Carlsberg Foundation.

**13.** Lorenz 1867 (German version), p. 243, heading note.

**14.** Lorenz 1867 (English version), p. 287, heading note.

**15.** Helge Kragh, «Ludvig Lorenz and His Non-Maxwellian Electrical Theory of Light», Physics in Perspective 20, no. 3, 221-253 (2018).

**16.** J. D. Jackson, «Examples of the zeroth theorem of the history of science», American Journal of Physics 76, 704-719 (2008), sec. II.D; arXiv:0708.4249.

**17.** Potter, arXiv:0811.2123.

**18.** Wong, arXiv:1012.4128.

**19.** Kragh, «Ludvig Lorenz (1867) on Light and Electricity», arXiv:1803.06371 (2018), introduction, on «Om Lyset», Tidsskrift for Physik og Chemi 6, 1-9 (1867).

**20.** Lorenz 1867 (English version), pp. 290 and 291; Danish version, p. 31.

**21.** Lorenz 1867 (English version), pp. 294-295; Danish version, p. 35.

**22.** Jackson, arXiv:0708.4249, sec. II.

**23.** Potter, arXiv:0811.2123, abstract.

**24.** Lorenz 1867 (English version), p. 299; Danish version, p. 41.

**25.** Lorenz 1867 (Danish version), p. 29.

**26.** Ibid., p. 32.

**27.** Ibid., p. 33.

**28.** Ibid., p. 34. In the English version, p. 294: «there is therefore some reason for taking $a = c/\sqrt{2}$».

**29.** Ibid., p. 33. English version, p. 293: «the equations (A) are transformed into the following differential equations».

**30.** Ibid. The work referred to is L. V. Lorenz, «Mémoire sur la théorie de l'élasticité des corps homogènes à élasticité constante», Journal für die reine und angewandte Mathematik 58, 329-351 (1861).

**31.** Ibid., p. 36. The work referred to is «Die Theorie des Lichtes, II», Annalen der Physik und Chemie 121, 579-600 (1864).

**32.** Ibid., p. 39. The work referred to is «Die Theorie des Lichtes, I», Annalen der Physik und Chemie 118, 111-145 (1863), p. 126. The reference with page is in the English version, p. 297.

**33.** Ibid. English version, p. 293: «These equations are satisfied, for instance, by».

**34.** Ibid., p. 42. English version, p. 300.

**35.** Jackson, arXiv:0708.4249, sec. II.D.

**36.** Wong, arXiv:1012.4128, abstract.

**37.** Potter, arXiv:0811.2123, appendix B.

**38.** Lorenz 1867 (Danish version), pp. 26-45. Count carried out on the complete text: two occurrences on p. 26 and one on p. 37.

**39.** Potter, arXiv:0810.1172, sec. 1.

**40.** Wong, arXiv:1012.4128.

**41.** Helge Kragh, «Ludvig Lorenz, Electromagnetism, and the Theory of Telephone Currents», arXiv:1606.00205 (2016), sec. 4.

**42.** Jackson, arXiv:0708.4249, sec. II.D.

**43.** Jackson and Okun, sec. II.C; and Bruce J. Hunt, The Maxwellians (Cornell University Press, Ithaca, 1991), pp. 115-118.

**44.** E. Wiechert, «Elektrodynamische Elementargesetze», in J. Bosscha (ed.), Recueil de travaux offerts par les auteurs à H. A. Lorentz, Archives néerlandaises des sciences exactes et naturelles, series 2, vol. 5, 549-573 (1900); also in Annalen der Physik, series 4, vol. 4, 667-689 (1901). The quotation, on p. 552.

**45.** Ibid., p. 553, equation (9).

**46.** Ibid., p. 553.

**47.** Jackson, arXiv:0708.4249, sec. II.D.

**48.** H. A. Lorentz, «Weiterbildung der Maxwellschen Theorie. Elektronentheorie», Encyklopädie der mathematischen Wissenschaften, Band V:2, Heft 1, V.14, 145-280 (1904), p. 157, equation (2). Dated December 1903.

**49.** Ibid., p. 157. The description agrees with that of Jackson and Okun, sec. II.E.

**50.** Ibid., p. 157, footnotes: A. Liénard, L'Éclairage électrique 16 (1898), pp. 5, 53, 106; and Wiechert, Arch. néerl. (2), 5 (1900), p. 549.

**51.** Ibid., p. 154, with footnotes to Gauss, Riemann, C. Neumann and Betti.

**52.** Kirk T. McDonald, «Helmholtz and the Velocity Gauge», Princeton University, 31 March 2018, updated 20 February 2019.

https://kirkmcd.princeton.edu/examples/helmholtz.pdf

**53.** H. Weyl, «Eine neue Erweiterung der Relativitätstheorie», Annalen der Physik 59, 101-133 (1919); and «Gravitation and the electron», Proceedings of the National Academy of Sciences 15, 323-334 (1929).

**54.** Jackson and Okun, sec. V.

**55.** Jackson and Okun, sec. IV.B.

**56.** W. Heitler, The Quantum Theory of Radiation, 2nd ed. (Oxford University Press, London, 1944), pp. 3-4.

**57.** J. C. Maxwell, «XXVI. On a Method of Making a Direct Comparison of Electrostatic with Electromagnetic Force; with a Note on the Electromagnetic Theory of Light», Philosophical Transactions of the Royal Society of London 158, 643-657 (1868); in The Scientific Papers of James Clerk Maxwell, ed. W. D. Niven (Cambridge University Press, 1890), vol. II, p. 137.

**58.** Jackson and Okun, sec. II.

**59.** Kirk T. McDonald, «Maxwell's Objection to Lorenz' Retarded Potentials», Princeton University, 2009, updated 2014.

https://kirkmcd.princeton.edu/examples/maxwell.pdf

**60.** Kragh, Physics in Perspective (2018). The review is by Gustav Radicke, Die Fortschritte der Physik 23, 197-200 (1870).

**61.** Kragh, Applied Optics 30, no. 33, 4688-4695 (1991); and Physics in Perspective (2018).

**62.** Jackson, arXiv:0708.4249, sec. VII.

**63.** Robert Nevels and Chang-Seok Shin, «Lorenz, Lorentz, and the gauge», IEEE Antennas and Propagation Magazine 43, no. 3, 70-71 (2001).

**64.** H. Hertz, «Ueber die Beziehungen zwischen den Maxwell'schen elektrodynamischen Grundgleichungen und den Grundgleichungen der gegnerischen Elektrodynamik», Annalen der Physik und Chemie 23, 84-103 (1884); authorized English translation by D. E. Jones and G. A. Schott in Miscellaneous Papers (Macmillan, London, 1896), pp. 273-290. The quotation, on p. 286.

**65.** Ibid., p. 276, footnote.

**66.** H. Hertz, «On the Relations between Light and Electricity», lecture at Heidelberg, 20 September 1889; in Miscellaneous Papers, pp. 313-327. The mention, on p. 318.